\documentclass[sigconf,nonacm,natbib=false]{acmart}
\AtBeginDocument{%
\providecommand\BibTeX{{%
    Bib\TeX}}}

\RequirePackage[
  datamodel=acmdatamodel,
  style=acmnumeric,
  ]{biblatex}

\usepackage{amssymb}
\usepackage{amsmath,amssymb,amsfonts}
\usepackage{algorithm}
\usepackage{algorithmic}
\usepackage{graphicx}
\usepackage{textcomp}
\usepackage{xcolor}
\usepackage{graphicx}
\usepackage{subcaption}
\usepackage{verbatim}
\usepackage{comment} 

\def\BibTeX{{\rm B\kern-.05em{\sc i\kern-.025em b}\kern-.08em
    T\kern-.1667em\lower.7ex\hbox{E}\kern-.125emX}}

\usepackage{xcolor}
\newboolean{showcomments}
\setboolean{showcomments}{false}
\ifthenelse{\boolean{showcomments}}
{ \newcommand{\mynote}[3]{
		\fbox{\bfseries\sffamily\scriptsize#1}
		{\small$\blacktriangleright$\textsf{\emph{\color{#3}{#2}}}$\blacktriangleleft$}}}
{ \newcommand{\mynote}[3]{}}
\newcommand{\shrink}[1]{}
\definecolor{pink}{rgb}{1,0.2,0.7}
\definecolor{purple}{rgb}{0.7,0,0.9}

\newcommand{\cl}[1]{\mynote{Chieh-Lin}{#1}{red}}
\newcommand{\gc}[1]{\mynote{Guan-Cheng}{#1}{green}}

\author{Guan-Cheng Chen}
\email{p96114214@gs.ncku.edu.tw}
\affiliation{%
  \institution{National Cheng Kung University}
  \city{Tainan}
  \country{Taiwan}
}

\author{Chieh-Lin Tsai}
\email{d09922013@csie.ntu.edu.tw}
\affiliation{%
  \institution{National Taiwan University}
  \city{Taipei}
  \country{Taiwan}
}

\author{Pei-Hsuan Tsai}
\email{phtsai@mail.ncku.edu.tw}
\affiliation{%
  \institution{National Cheng Kung University}
  \city{Tainan}
  \country{Taiwan}
}

\author{Yuan-Hao Chang}
\email{johnson@csie.ntu.edu.tw}
\affiliation{%
  \institution{National Taiwan University}
  \city{Taipei}
  \country{Taiwan}
}

\begin{document}

\title{Sensitivity-Aware Mixed-Precision Quantization for ReRAM-based Computing-in-Memory}

\begin{abstract}
Compute-In-Memory (CIM) systems, particularly those utilizing ReRAM and memristive technologies, offer a promising path toward energy-efficient neural network computation. However, conventional quantization and compression techniques often fail to fully optimize performance and efficiency in these architectures. In this work, we present a structured quantization method that combines sensitivity analysis with mixed-precision strategies to enhance weight storage and computational performance on ReRAM-based CIM systems. Our approach improves ReRAM Crossbar utilization, significantly reducing power consumption, latency, and computational load, while maintaining high accuracy. Experimental results show 86.33\% accuracy at 70\% compression, alongside a 40\% reduction in power consumption, demonstrating the method's effectiveness for power-constrained applications.
\end{abstract}

\maketitle

\section{Introduction}
Deep learning has revolutionized fields such as computer vision, natural language processing (NLP), and robotics. However, deploying complex models in resource-constrained environments, such as wearable devices, smart sensors, and industrial automation systems, remains a significant challenge. These platforms are often subject to strict limitations in computational resources and energy availability, which negatively impact inference speed and accuracy, particularly in latency-sensitive tasks like image recognition and voice interaction.

Computing-in-Memory (CIM) architectures have emerged as a promising solution to these constraints by addressing the memory bottlenecks inherent in traditional von Neumann architectures. By performing computations directly within memory arrays in an analog manner, CIM systems significantly reduce data movement, leading to improved latency and energy efficiency. This analog computation paradigm enables massively parallel operations such as matrix-vector multiplication, making CIM particularly suitable for accelerating deep neural network workloads in low-power environments. Notable advancements in this field include ISAAC\cite{7551379} and PRIME\cite{7551380}, which utilize ReRAM-based crossbar arrays to realize efficient in-memory analog computation for deep learning inference.

Despite these architectural advancements, a critical challenge remains. Most optimization techniques for deep neural networks are designed for conventional digital systems. Consequently, compression methods such as pruning and quantization, which are effective for reducing computational overhead in digital platforms and are well-suited to resource-constrained environments, often fail to deliver comparable benefits in analog CIM systems. This limitation arises from analog circuit non-idealities, data mapping constraints, and sensitivity to sparsity patterns, all of which can reduce crossbar utilization and lead to suboptimal energy efficiency and accuracy.

To tackle these challenges, researchers have developed specialized quantization and pruning techniques specifically designed for CIM (Computing-in-Memory) architectures. \textit{Quantization} reduces bit-width, thereby lowering computational and storage complexity to improve energy efficiency. Genetic Algorithm-Based Quantization (GAQ) \cite{9218737} dynamically adjusts layer-wise precision to balance accuracy and hardware cost, while Regularization with Quantization Training (RQT) \cite{9045192} integrates quantization into the training process to enhance robustness. \textit{Pruning} techniques, on the other hand, exploit sparsity to skip less important weights and minimize redundant computations. Structured pruning, a form of sparsity, reduces computation overhead while maintaining accuracy, resulting in improved hardware performance. Fine-Grained DCNN Pruning \cite{9218523} reduces resource usage by removing non-critical rows and columns in ReRAM crossbars, and Multi-Group Lasso Pruning \cite{9387391} enhances structural sparsity by aligning column-group pruning with 4-bit quantization.

In addition, recent research on CIM systems, particularly those based on ReRAM and memristors technologies, has explored the integration of pruning and quantization to further enhance resource utilization and reduce power consumption. Mixed-precision techniques have emerged as a promising direction, where the bit-width of weights is dynamically adjusted—assigning higher precision to critical weights and lower precision to less sensitive ones. This selective quantization approach effectively balances computational load and energy efficiency, making it especially suitable for power-constrained environments.

For example, \cite{10075408} proposed a NAS-based bit-level sparsity-tolerant method to minimize performance loss under low-precision quantization and sparse training  but
it is computationally expensive, making it impractical for resource-limited environments. Meanwhile, \cite{9852786} introduced an AutoML-based mixed-precision framework using differentiable architecture search to adjust CNN bit-width for fine-grained quantization in ReRam arrays. \cite{9218724} developed an automated quantization and mapping framework for general-purpose ReRam accelerators using a two-stage learning strategy. Finally, \cite{9425549} proposed a dynamic mixed-precision method using Hessian matrix analysis to adjust quantization precision based on weight sensitivity. \cl{Drawback of 8 and 9? Is HAP outperforms 8 and 9?}\gc{Papers 8 and 9 use NAS, which requires extensive computational resources. Comparing with them may be unfair, as our advantage lies in generating results through computation rather than the near-exhaustive search of NAS. Paper 10 is a more suitable comparison.}

\cl{We focus on HAP, seems HAP is our main comparing opponent}
\gc{We chose HAP because it offers good replicability. Although several papers related to Paper 10 could serve as comparisons , their experiments cannot be replicated. I have also switched to using the HAWQ v2 code, which implements the same algorithm as HAP. The only difference is that HAP is used for pruning, while HAWQ is applied to quantization, making it a better match for our work.}
Although Hessian-driven mixed-precision methods offer faster processing, they typically rely on unstructured quantization schemes. This leads to significant inefficiencies in ReRAM-based CIM systems, as the resulting misalignment causes many memory cells to be underutilized. Moreover, effective clustering of weights based on precision is essential for maximizing crossbar utilization but is frequently neglected, resulting in suboptimal hardware resource usage. Additionally, the absence of a unified processing pipeline for handling mixed-precision outputs introduces further misalignment, which can degrade model accuracy and reduce overall system performance. These limitations highlight the need for further research to develop more structured and hardware-aware quantization strategies.

In this work, we present a \textit{sensitivity-aware structured mixed-precision quantization framework} designed to overcome the limitations of conventional compression techniques when applied to ReRAM-based Computing-in-Memory (CIM) systems. Conventional mix-precision methods, such as Hessian-Aware Pruning (HAP), often underperform in CIM architectures due to incompatibility with column-wise weight storage and reliance on unstructured sparsity. These limitations lead to inefficient crossbar utilization, increased energy consumption, and degraded computational efficiency.

To address these challenges, our framework leverages sensitivity analysis to guide the initial grouping of weights into high- and low-sensitivity clusters, each assigned an appropriate bit-width. This \textit{tiered quantization strategy}, inspired by the Optimal Brain Damage (OBD) technique~\cite{NIPS1989_6c9882bb}, uses the Hessian matrix to evaluate weight importance and align quantization precision accordingly.

Furthermore, we introduce a \textit{dynamic clustering mechanism} \cl{There are two cluster here?} \gc{Yes — there are two clusters, typically configured as 8/4-bit and 4/2-bit. Co-placing crossbars with different bit widths enables mixed-precision operation.}that refines these sensitivity-based clusters to match crossbar constraints. This mechanism ensures that mixed-precision weights are efficiently mapped to the hardware, enabling independent processing and subsequent aggregation. It improves crossbar utilization and mitigates inefficiencies arising from unstructured quantization schemes.

Experimental results demonstrate the effectiveness of our proposed techniques. Compared to prior methods, our framework achieves \cl{Need to update the result} 84.63\% accuracy (vs. 74.8\%), reduces ReRAM latency by 65\%, and cuts power consumption by 70\%. ADC read latency and energy are also reduced by 30\% and 60\%, respectively. At a 70\% compression ratio, our method sustains 86.33\% accuracy with a 40\% reduction in power consumption, validating its applicability to power-constrained environments.

\noindent To conclude, this work makes the following key contributions:
\begin{itemize}
    \item We propose a \textit{sensitivity-aware structured mixed-precision quantization method} tailored for analog ReRAM-based CIM systems. This approach addresses the inefficiencies of conventional compression techniques by improving weight-to-crossbar mapping through structured, sensitivity-guided quantization.
    \item We introduce a \textit{dynamic clustering mechanism} \cl{Should use "Dynamic Threshold Optimization" instead} \gc{Great, that’s an excellent suggestion.}for mixed-precision quantization that enables independent computation of high- and low-precision clusters, guided by Hessian-based sensitivity analysis.
    \item We provide a comprehensive evaluation using ReRAM-based crossbar simulations, validating our method's effectiveness in optimizing performance, energy efficiency, and accuracy for CIM accelerators.
\end{itemize}

\section{Background and Related Work}
Resistive Random Access Memory (ReRAM) is a non-volatile memory technology that stores data by modulating the resistance of materials, maintaining information even without power. Its compact design, typically based on a crossbar structure, allows for high scalability and low power consumption. ReRAM devices switch between high and low resistance states through the formation and dissolution of conductive filaments, making them ideal for high-density storage applications and deep learning accelerators\cite{PAN20141}\cite{ZHANG2014415}\cite{6835201}\cite{7312469}. 

\subsection{ReRAM Crossbar in Deep Learning Accelerators}
In Computing-In-Memory (CIM) architectures, ReRAM crossbars enable efficient matrix-vector multiplication (MVM). Each ReRAM cell represents matrix elements as conductance, and input vectors are applied as voltages. The resulting currents, representing the dot product, are accumulated, allowing for parallel computations with \( O(1) \) time complexity\cite{Sze2020-be}.

\begin{figure}[htbp]
\centerline{\includegraphics[scale=0.40]{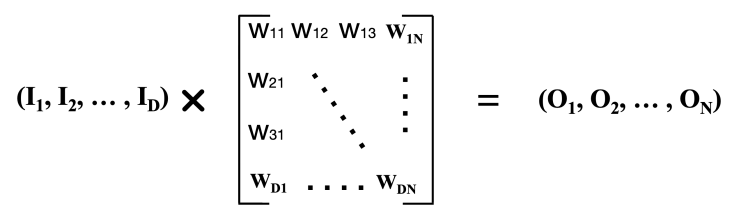}}
\caption{The crossbar array mathematically implements matrix-vector multiplication. Input voltages \(I_1, I_2, \dots, I_D\) are applied across conductance values \(W_{ij}\), generating output currents \(O_j = \sum_{i=1}^{D} I_i W_{ij}\) through parallel computation.}
\label{fig}
\end{figure}

For convolutional neural networks (CNNs), the \( K \times K \) convolution kernel is split into \( 1 \times 1 \times D \) vectors and processed in parallel across the ReRAM crossbar. Sub-matrix computations are accumulated to form the final output feature maps (OFMs), maximizing parallelism, reducing energy consumption, and enhancing performance\cite{8702715}\cite{Gopalakrishnan2020-of}.

\begin{figure}[htbp]
\centerline{\includegraphics[scale=0.33]{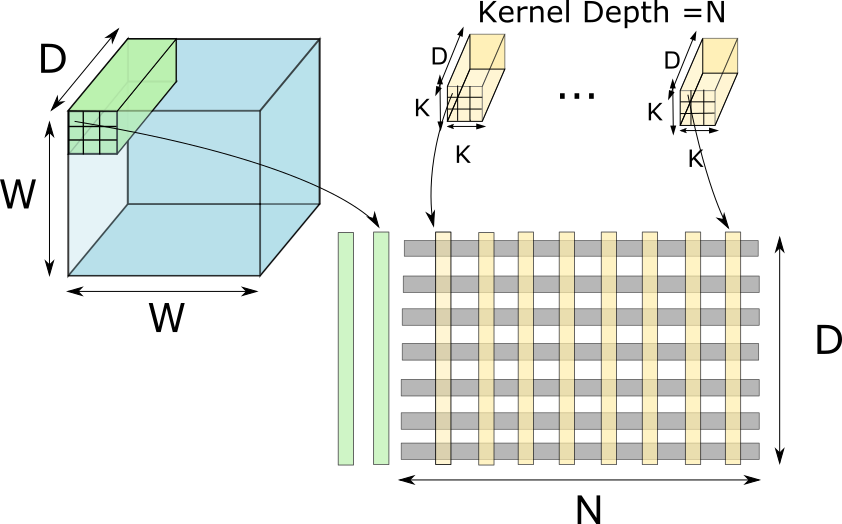}}
\caption{CNN \(K \times K\) kernels are split and processed in parallel on ReRAM crossbars for efficient OFMs.}
\label{fig}
\end{figure}

\subsection{Sparsity and ADC Resolution in ReRAM-Based Systems}
Sparse matrix operations and ADC resolution are critical factors in the energy efficiency of ReRAM-based systems. Pruning and quantization reduce neural network computational load but create challenges in ReRAM architectures. Unstructured sparsity leads to inefficiencies in crossbar arrays, as even inactive rows or columns must be computed during matrix-vector multiplication, causing unnecessary power consumption and reducing hardware efficiency\cite{9855854}\cite{9965422}\cite{10.1109/TPAMI.2022.3195774}.

\begin{figure}[htbp]
    \centering
    \begin{subfigure}[b]{0.15\textwidth}
        \centering
        \includegraphics[width=\textwidth]{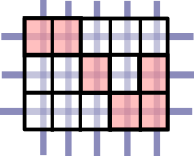}
        \caption{}
        \label{fig:subfig_a}
    \end{subfigure}
    \hfill
    \begin{subfigure}[b]{0.15\textwidth}
        \centering
        \includegraphics[width=\textwidth]{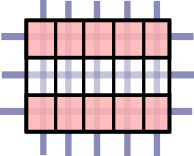}
        \caption{}        
        \label{fig:subfig_b}
    \end{subfigure}
    \hfill
    \begin{subfigure}[b]{0.15\textwidth}
        \centering
        \includegraphics[width=\textwidth]{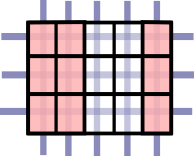}
        \caption{}
        \label{fig:subfig_c}
    \end{subfigure}
    \caption{Sparsity in ReRAM crossbars: (a) unstructured sparsity causes inefficiency, (b) structured pruning enhances efficiency, (c) optimal pruning eliminates inactive rows/columns, maximizing crossbar energy efficiency and performance.}
    \label{fig:three_in_a_row}
\end{figure}

Similarly, ADC resolution significantly impacts energy use. Higher resolutions increase power consumption exponentially, while reducing ADC resolution by just one bit can improve computational efficiency by 50\% and boost energy efficiency by 87\%\cite{7551379}\cite{10.1145/3489517.3530476}. Thus, optimizing ADC resolution and addressing sparsity issues are key to enhancing the overall performance and energy efficiency of ReRAM-based CNN accelerators.

\subsection{Weight Sensitivity in Neural Networks}
Weight sensitivity in neural networks refers to how changes in individual weights affect model performance, particularly prediction accuracy. By analyzing this, we can identify critical weights and determine which can be pruned or adjusted with minimal impact on performance.

Weight sensitivity is quantified using second-order derivatives of the error function, represented by the Hessian matrix \( H \)\cite{NIPS1989_6c9882bb}. The error function \( E(w) \) can be approximated by a Taylor series expansion around a local minimum as:

\[
\Delta E = \nabla E(w)^T \Delta w + \frac{1}{2} \Delta w^T H \Delta w + O(\|\Delta w\|^3)
\]

Where:
\begin{enumerate}
    \item \( \nabla E(w)^T \Delta w \) is the first-order term (gradient term), representing the change in error due to the linear approximation.
    \item \( \frac{1}{2} \Delta w^T H \Delta w \) is the second-order term, representing the curvature of the error function, where \( H \) is the Hessian matrix (the matrix of second-order derivatives).
    \item \( O(\|\Delta w\|^3) \) is the third-order term, representing higher-order changes that are typically neglected for small perturbations in \( w \).
\end{enumerate}

Since the network is assumed to be at a local minimum, the first-order derivative (gradient) vanishes, leaving the second-order approximation for practical purposes   .

However, calculating the full Hessian matrix in neural networks is computationally expensive, particularly for deep models. Diagonal approximation in Optimal Brain Damage prunes weights using Hessian diagonal elements for efficiency\cite{NIPS1989_6c9882bb}. Optimal Brain Surgeon prunes weights using full Hessian via Woodbury identity for precise optimization\cite{298572}. Hutchinson's Algorithm approximates Hessian trace efficiently, aiding mixed-precision quantization by averaging eigenvalues\cite{dong2019hawq}\cite{dong2020hawq}. 
The Eigen Damage (ED) baseline method extends OBD and OBS to structured pruning, summarizing parameter changes at the filter level, enhancing granularity\cite{wang2019eigendamage}. \cl{Which one do we use?}

\cl{HAP is the SOTA? What is its weakness?}

In particular, Hessian-Aware Pruning (HAP)\cite{yu2022hessian} represents the culmination of these techniques, efficiently pruning neural networks by using an approximation based on the Hessian trace. The loss perturbation caused by pruning is approximated as:

\[
\Delta L \approx \frac{1}{2} w_p^T \frac{\text{Trace}(H_{p,p})}{p} w_p
\]

HAP uses Hutchinson’s method to estimate the Hessian trace, which involves applying random vectors \( v_i \) to compute:

\[
\text{Trace}(H) \approx \frac{1}{m} \sum_{i=1}^{m} v_i^T H v_i
\]

This allows efficient second-order pruning without explicitly computing the full Hessian matrix.

\subsection{Robustness of Neural Networks}
Robustness in neural networks refers to the ability of a model to maintain performance when subjected to variations in input data or perturbations in its structure. From the geometric and statistical perspectives, robustness is closely linked to the network's loss landscape and the Fisher Information Matrix (FIM). The FIM provides a measure of how sensitive the output of a neural network is to changes in its parameters, and its eigenvalues can indicate the local flatness or sharpness of the loss landscape. Networks with flatter minima, characterized by smaller FIM eigenvalues, tend to generalize better because they are less sensitive to small parameter perturbations. On the other hand, sharp minima, associated with larger FIM eigenvalues, can lead to worse generalization due to overfitting, where the model becomes highly sensitive to small changes in the input or parameters\cite{karakida2019universal}\cite{liang2019fisher}.

In addition, visualizing the loss landscape of neural networks has shown that models with flatter minima are generally more robust. Factors such as skip connections and batch normalization can significantly influence the sharpness or flatness of the minima found during training. These factors allow the network to avoid chaotic loss landscapes, promoting better generalization, and thus improving robustness\cite{li2018visualizing}.

\begin{figure*}[ht]
    \centering
    \includegraphics[width=\textwidth]{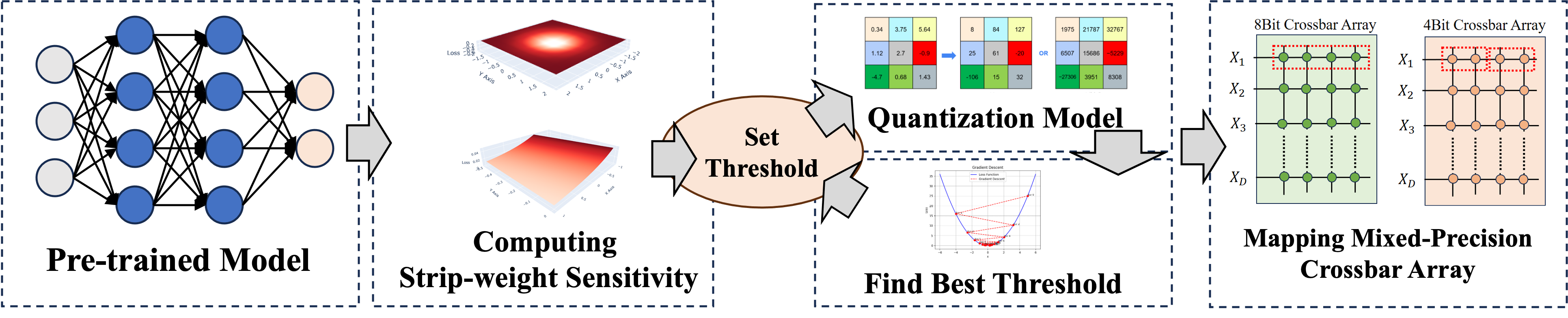}
    \caption{The process of adaptive quantization for deep learning models on ReRAM Crossbar architectures. Starting with a pre-trained model, the strip-weight sensitivity is computed using the Hessian matrix to assess the impact of each weight on model accuracy. A threshold is set to categorize high and low-sensitivity weights, which undergo different levels of quantization. The optimal threshold is determined through iterative optimization. Finally, the quantized weights are mapped onto Crossbar arrays, with high-precision weights assigned to 8-bit Crossbars and low-precision ones to 4-bit, optimizing both performance and energy efficiency.}
    \label{fig:pruning_types}
\end{figure*}

\section{Motivation}

\cl{Need a more comprehensive motivation, maybe an illustration about why other works can't fully utilize the Xbar, and how we solve this issue}

This work is motivated by the urgent need to enhance the operational efficiency and energy savings of deep neural networks on ReRAM crossbar architectures, particularly in low-power environments like IoT devices.

In contrast to previous mixed-precision methods that often sacrifice either performance or energy efficiency on Computing-In-Memory (CIM) architectures, we are interested in developing a structured mixed-precision quantization method tailored for CIM architectures. Our objective is to optimize both computational efficiency and energy consumption, allowing CNNs to perform effectively within low-power environments while preserving model accuracy.

The technical problem arises from the need to balance irregular quantization-induced sparsity with ReRAM’s crossbar architecture, where hardware underutilization can negate the advantages of quantization. At the same time, we must ensure that the mixed-precision quantization of weights and inputs aligns with hardware constraints to maintain both accuracy and efficiency.

To achieve this, the following factors must be considered:
\begin{enumerate}
    \item Efficient Mapping of Quantized Weights: Ensuring ReRAM crossbar resources are fully utilized by aligning quantization strategies with the underlying hardware architecture.
    \item Crossbar Size and Weight Allocation: Properly matching crossbar size with the precision allocation of weights to avoid resource waste while preserving model accuracy.
    \item Consistency in Data Processing: Handling data type conversion and layout consistency during mixed-precision computations to guarantee accuracy in low-power environments.
\end{enumerate}

In summary, our motivation is to address these challenges and develop a comprehensive solution that optimizes both energy efficiency and model performance in resource-constrained CIM systems.

\section{Methodology}
This study presents three solutions to address the challenges of deep learning model compression in ReRAM crossbar architectures. These solutions focus on optimizing crossbar design, implementing mixed-precision computation, and efficiently managing clustered weights.

First, structured quantization combined with sensitivity-based clustering improves computation efficiency by organizing weights based on their sensitivity to quantization. This reduces computational load while maintaining the parallel processing power of the crossbar, ensuring minimal resource waste and accurate results. Next, dynamic clustering for weight mapping adjusts cluster boundaries to match the size of the crossbar, maximizing hardware utilization and preventing performance degradation from uneven weight distribution. Lastly, a precision-aware parallel computation mechanism is introduced to manage mixed-precision operations efficiently. This approach aligns data with different precisions for parallel computation, minimizing errors and enhancing overall system performance.

These solutions target key technical challenges related to quantization, and weight deployment, offering a comprehensive strategy for optimizing ReRAM-based deep learning accelerators.

\subsection{Structured Quantization with Sensitivity Clustering}\label{AA}
In this study, we analyzed the weight structures within convolutional neural networks (CNNs) and introduced the concept of strip-weight as the fundamental computational unit. The use of strip-weight offers a significant advantage in addressing the problem of irregular sparsity caused by traditional quantization techniques, which often leads to inefficient utilization of hardware resources in ReRAM crossbar architectures. 

\begin{figure}[htbp]
\centerline{\includegraphics[scale=0.22]{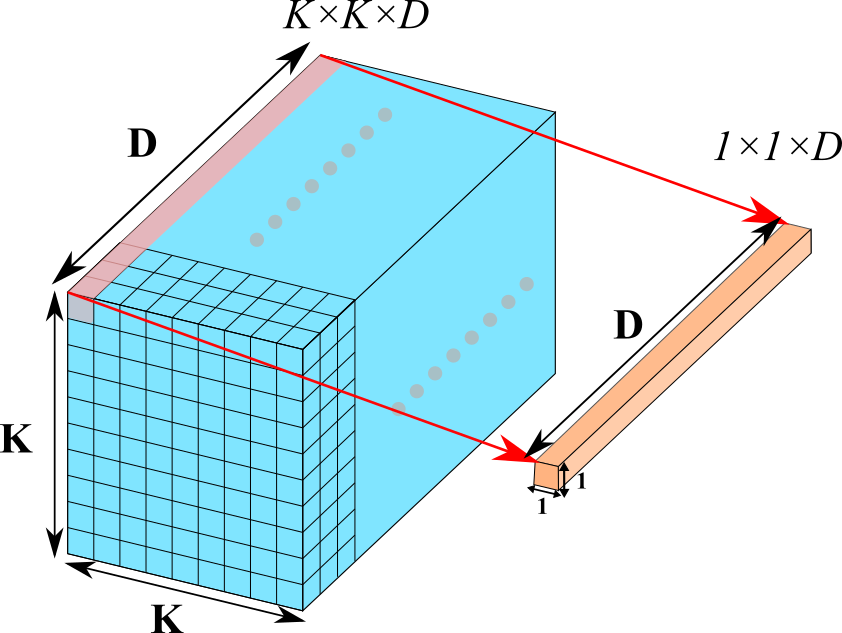}}
\caption{CNN weight structure: Strip-weight units serve as the fundamental computational blocks in convolution operations.}
\label{fig}
\end{figure}

By dividing the convolutional kernel into smaller \(1 \times 1 \times D\) units, the strip-weight ensures that the ReRAM crossbar resources are fully utilized. This method preserves the structural integrity of the pruned weights, preventing hardware performance degradation and optimizing the parallel computation capabilities of the crossbar array.

In a convolutional layer, the activation output \( z \) is obtained by performing a convolution between the weights \( w \) and the input feature map \( x \). When weights are redefined as strip-weights, the convolution operation can be represented as:

\[
z^l_{ij} = \sum_{m=1}^{M} \sum_{n=1}^{N} w^l_{mn} \cdot x_{(i+m-1)(j+n-1)}
\]

where:
\begin{itemize}
    \item \( w^l_{mn} \) is a \( 1 \times 1 \times D \) strip-weight vector for the \( l \)-th layer.
    \item \( x_{(i+m-1)(j+n-1)} \) is the corresponding portion of the input feature map, also of size \( 1 \times 1 \times D \).
\end{itemize}

Here, each \( w^l_{mn} \) represents a strip-weight within the convolutional kernel, which preserves the structure and ensures that all ReRAM crossbar resources are fully utilized \cl{How to fully utilize crossbar?}. The input feature map \( x \) is sampled according to the kernel's sliding window, matching the dimensions of each strip-weight. This structure allows for more efficient computations by mapping smaller and more manageable units of weights onto the crossbar array.

We assess the sensitivity of each strip-weight using the Hessian matrix, which reflects how weights impact model accuracy. Higher sensitivity means greater importance during quantization.

Fully computing the Hessian in large networks is challenging. To simplify, HAP \cl{Refer to previous work? or we propose this?} showed that focusing on the diagonal elements of the Hessian is sufficient, as the off-diagonal elements contribute little. Using the trace of the Hessian, we approximate sensitivity as:

\[
\Delta L \approx \frac{\text{Trace}(H_{p,p})}{2p} \|w_p\|_2^2
\]

This method helps prune weights with minimal accuracy impact, reducing computational complexity while preserving performance.

In our framework, we represent convolutional layer weights using strip-weights, defined as \(1 \times 1 \times D\) weight vectors, where \(D\) is the depth of the convolutional kernel. A convolutional layer with dimensions \(K \times K \times D \times N\) is thus redefined into \(K \times K \times N\) strip-weights, where \(N\) is the number of output channels. We focus on calculating the sensitivity of each strip-weight by analyzing its Hessian matrix to assess its importance.

The Hessian matrix for a strip-weight, \(H_{\text{strip}}\), measures the second-order partial derivatives of the activation output \(z_{\text{ijn}}^{\text{strip}}\) with respect to the strip-weight. This helps to quantify each strip-weight's influence on the model:

\[
H_{\text{strip}} = \frac{1}{n} \sum_{j=1}^{n} \sum_{i=1}^{K^2 \times N} \sum_{jn} \frac{\partial z_{\text{ijn}}^{\text{strip}}}{\partial[w_1, \dots, w_{K^2N}]}
\]

where:
\begin{itemize}
    \item \(H_{\text{strip}}\) is the Hessian matrix for the strip-weight,
    \item \(n\) is the number of training samples,
    \item \(K^2 \times N\) is the total number of strip-weights in the layer,
    \item \(z_{\text{ijn}}^{\text{strip}}\) represents the activation output corresponding to the \(i\)-th position and the \(jn\)-th strip-weight.
\end{itemize}

The sensitivity score is derived from the trace of the Hessian matrix:

\[
\text{Sensitivity Score}_{\text{Strip}} \approx \frac{\text{Trace}(H_{\text{strip}})}{2p_{\text{strip}}} \| w_{\text{strip}} \|_2^2
\]

This score reflects the strip-weight's overall impact on the model, with the trace of \(H_{\text{strip}}\) indicating its importance. Strip-weights with higher sensitivity should retain higher precision during quantization. Here, \(p_{\text{strip}}\) is the number of weights in each strip-weight, and \(\| w_{\text{strip}} \|_2^2\) is the L2 norm, representing the weight magnitude.

For each strip-weight \( \phi_i \), we calculate its sensitivity \( s_i \) using the previously mentioned formula. To balance computational efficiency and accuracy in CNNs, we employ a sensitivity-based clustering strategy.

Strip-weights are first sorted by their sensitivities \( s_i \), and a threshold \( T \) is set to distinguish high-sensitivity from low-sensitivity weights. Strip-weights with \( s_i > T \) are considered critical for accuracy, so they are assigned to high-bit (e.g., 8-bit) ReRAM crossbars, ensuring precision and minimizing accuracy loss during quantization. Conversely, strip-weights with \( s_i \leq T \) have less impact on accuracy and are assigned to low-bit (e.g., 4-bit) crossbars, which reduces precision requirements and power consumption without significantly affecting performance.

This approach optimizes resource allocation by clustering weights as follows:

\[
\text{cluster}(\phi_i) =
\begin{cases} 
\text{High Precision cluster}, & \text{if } s_i > T \\
\text{Low Precision cluster}, & \text{if } s_i \leq T
\end{cases}
\]

By tailoring precision based on sensitivity, this strategy maximizes computational efficiency in ReRAM crossbars, effectively balancing accuracy and resource usage while addressing irregular sparsity.

\subsection{Dynamic Clustering for Weight Mapping}
Setting the threshold \( T \) for sensitivity-based clustering presents two main challenges:

\begin{enumerate}
    \item Threshold Selection: If \( T \) is too high, power consumption decreases but model accuracy suffers. If too low, most strip-weights are assigned to high-bit crossbars, increasing power consumption. Thus, \( T \) must be experimentally tuned to balance accuracy and energy efficiency based on the model’s sensitivity distribution.
    
    \item Weight Mapping: Efficient mapping requires strip-weights to match the crossbar's capacity. A mismatch between high-bit and low-bit strip-weights forces the activation of extra crossbars, reducing computational efficiency and increasing power use.
\end{enumerate}

To ensure efficient computation in the ReRAM crossbar, we developed a dynamic clustering method to address the issues of threshold setting and weight mapping.

We propose a threshold-finding algorithm based on minimizing the difference in the Fisher Information Matrix (FIM) before and after compression. The FIM quantifies uncertainty in estimating model parameters and reflects the model’s robustness and generalization ability. By minimizing the difference in FIM between the original and compressed models, we retain essential statistical properties while reducing model complexity.

The FIM is defined as:

\[
F = E \left[ \left( \frac{\partial \log p(x; \theta)}{\partial \theta} \right) \left( \frac{\partial \log p(x; \theta)}{\partial \theta} \right)^T \right]
\]

where:
\begin{itemize}
    \item \(p(x; \theta)\) is the model’s probability distribution,
    \item \(\theta\) represents the model parameters (weights),
    \item \(E\) is the expectation based on the data distribution.
\end{itemize}

This approach finds an optimal compression threshold \( T \) by minimizing the difference in FIM between the original model \( \theta \) and the compressed model \( \theta_c \), thus balancing compression with maintaining performance.

\begin{algorithm}[h]
\caption{Optimal Compression Threshold via Minimizing Fisher Information Difference}
\begin{algorithmic}[1]
\STATE \textbf{Input:} Model parameters $\theta$, initial threshold $T_0$, learning rate $\eta$, tolerance $\epsilon$, max iterations $N$
\STATE \textbf{Output:} Optimal threshold $T$
\STATE Initialize $T \gets T_0$
\STATE Compute Fisher Information: $F_0 \gets \text{FIM}(\theta)$
\FOR{$k = 1$ to $N$}
    \STATE Compress: $\theta_c \gets \text{Compress}(\theta, T)$
    \STATE Compute $F \gets \text{FIM}(\theta_c)$
    \STATE Loss: $L \gets \| F - F_0 \|_F^2$
    \STATE Gradient: $g \gets 2 \text{Tr}\left( (F - F_0) \frac{\partial F}{\partial T} \right)$
    \STATE Update $T \gets T - \eta g$
    \IF{$\| F - F_0 \|_F \leq \epsilon$}
        \STATE \textbf{break}
    \ENDIF
\ENDFOR
\STATE \textbf{Return:} $T$
\end{algorithmic}
\end{algorithm}

The compression algorithm involves several steps. First, we calculate the Fisher Information Matrix (FIM) for the original model, starting with an initial threshold \( T_0 = 1 \), representing maximum compression, where all strip-weights are assigned low-bit precision. This serves as the reference for the compression process.

In each iteration, we apply a threshold \( T \) to cluster the weights using Structured Quantization with Sensitivity Clustering. Strip-weights above \( T \) are assigned high-bit precision (e.g., 8-bit), while those below\( T \) use low-bit precision (e.g., 4-bit). This reduces computational overhead while preserving model performance.

The compressed model is denoted \( \theta_c \), represented as:

\[
\theta_c = \text{Compress}(\theta, T)
\]

After compression, we recalculate the FIM and measure the difference between the original and compressed models using the Frobenius norm:

\[
L = \| F - F_0 \|_F^2
\]

We minimize this loss using gradient descent, adjusting the threshold \( T \) based on the gradient:

\[
g = 2 \, \text{Tr} \left( (F - F_0) \frac{\partial F}{\partial T} \right)
\]

Updating the threshold with:

\[
T= T - \eta g
\]

This process repeats until the difference in FIMs is below a tolerance \( \epsilon \), ensuring the compressed model closely matches the original in terms of statistical properties. The final threshold \( T' \) is used to assign high-sensitivity strip-weights to high-bit crossbars and low-sensitivity ones to low-bit crossbars, optimizing energy efficiency while maintaining accuracy.

To optimize the mapping of weights onto the ReRAM crossbar, we introduce a dynamic adjustment strategy for strip-weights prior to each mapping. This ensures efficient utilization of crossbar resources. In cases where the number of high-bit strip-weights \( q \) is not divisible by the crossbar capacity \( C \), the threshold \( T \) is incrementally adjusted to reduce \( q \) and make it a multiple of \( C \). This reallocation optimizes the use of high-bit crossbars, maximizing computational efficiency without significantly impacting power consumption due to low-bit operations.

The threshold \( T \) is recalculated before each mapping to ensure that the high-bit strip-weights efficiently fill the crossbar rows. While this may slightly reduce low-bit crossbar utilization, the low power consumption of low-bit calculations ensures that overall computational efficiency improves significantly.

Our proposed dynamic clustering method begins by quantifying the model’s generalization ability through the Fisher Information Matrix (FIM) and uses gradient descent to approach the optimal threshold \( T' \). Following this, we further adjust \( T \) dynamically to optimize weight mapping onto the crossbar, ensuring efficient computational resource utilization, while maintaining model accuracy and reducing power consumption.

This dynamic adjustment mechanism enables efficient weight mapping, enhances crossbar resource usage, and ensures an optimal balance between energy efficiency and performance during inference.

\subsection{Precision-Coordinated Parallel Computation}
In this method, we optimized the operation of convolutional neural networks (CNNs) on ReRAM crossbars by addressing alignment errors in mixed-precision systems. We restructured the traditional \( K \times K \) convolution kernel into strip-weights of \( 1 \times 1 \times D \), splitting them into high-bit and low-bit sections for parallel processing. A \( K \times K \times N \) convolution kernel is divided into \( q \) high-bit and \( p \) low-bit strip-weights, ensuring \( q + p = R \), where \( R \) is the total number of strip-weights.

\begin{figure}[htbp]
\centerline{\includegraphics[scale=0.27]{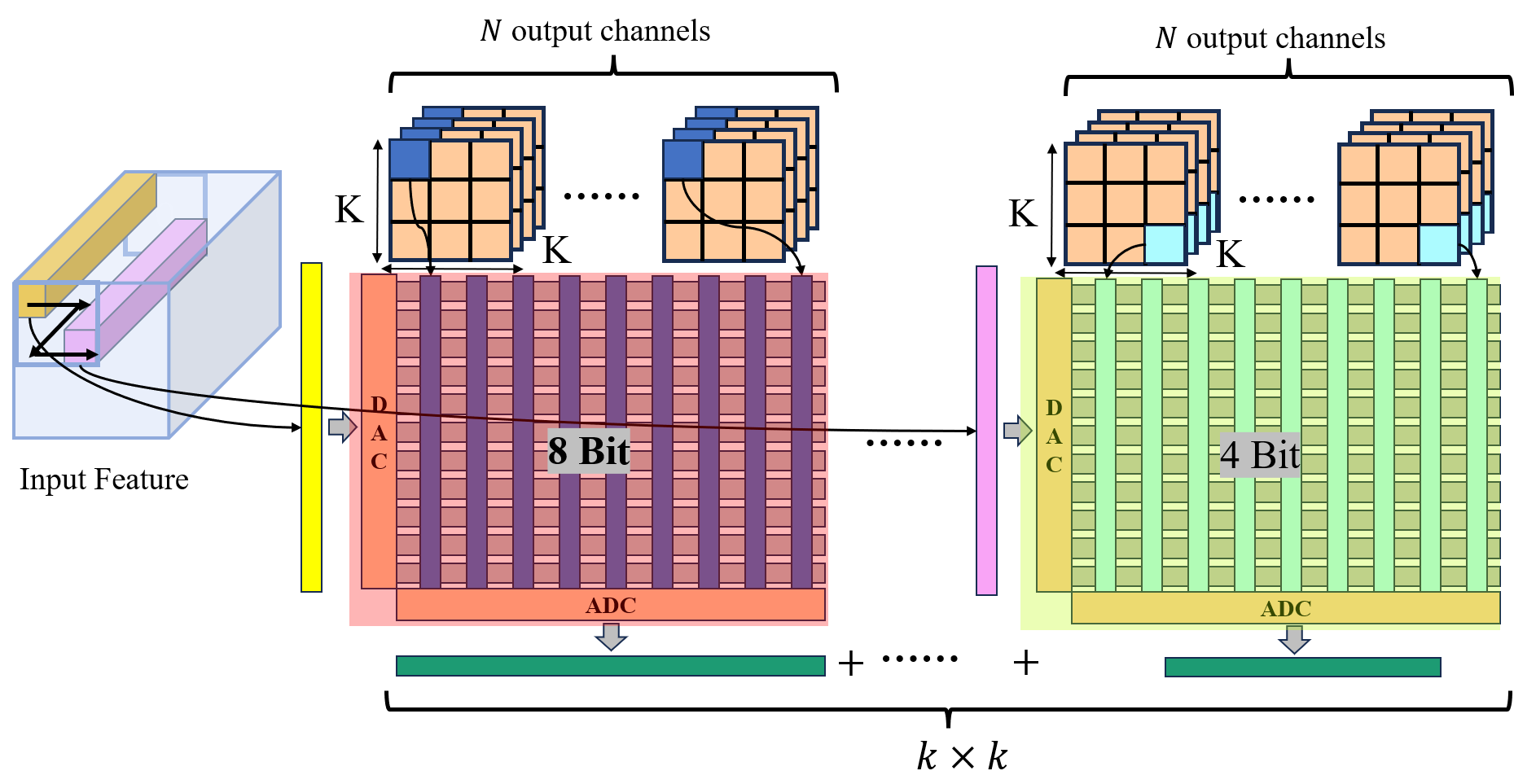}}
\caption{Convolution kernel split into high-bit and low-bit strip-weights, processed on 8-bit and 4-bit crossbars.}
\label{fig}
\end{figure}

The convolution is computed as \( Z = W \times A \), where \( W \) is split into high-bit \( W_q \) and low-bit \( W_p \). These are processed separately: \( Z_q = W_q \times A \) on an 8-bit crossbar and \( Z_p = W_p \times A \) on a 4-bit crossbar. To avoid alignment errors when summing these results, we apply a stepwise accumulation. The low-bit result \( Z_p \) is converted to match the high-bit format before final accumulation:

\[
Z = Z_q + \text{expand}(Z_p)
\]

\begin{figure}[htbp]
\centerline{\includegraphics[scale=0.7]{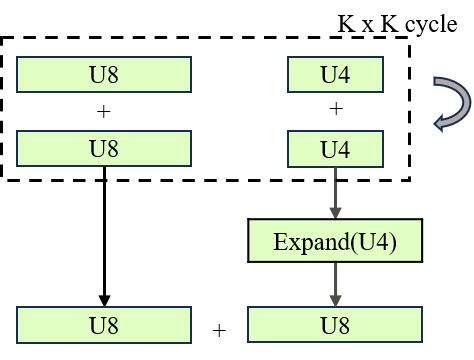}}
\caption{Low-bit results are expanded to match high-bit format for accurate stepwise accumulation in convolution.}
\label{fig}
\end{figure}

This method ensures that the low-bit result is aligned with the high-bit result, preventing errors in mixed-precision accumulation and fully utilizing the ReRAM crossbar's parallel capabilities. It significantly reduces resource consumption, improves energy efficiency, and maintains model accuracy in CNNs.

\section{Experiment}
We assessed our mixed-precision computation model on CIFAR-10 with a ResNet backbone and compared it to Hessian-Aware Pruning (HAP)\cite{hap_github} using its open-source code. Evaluations were performed in DNN+NeuroSim\cite{chen2017neurosim+}\cite{peng2019dnn+}\cite{chen2018neurosim}, which pairs with PyTorch to simulate ReRAM-crossbar latency, energy, and ADC-precision effects. While both methods follow standard CIFAR-10/ResNet protocols, each was run under its customary checkpoints and simulator settings, with mixed precision applied where it is methodologically supported. Under these conditions, our method achieves higher post-compression accuracy, reduced latency via better crossbar utilization, and lower power—particularly in mixed-precision configurations—demonstrating strong efficiency for CIM architectures.

Under a strip-wise mixed-precision pipeline, each convolutional layer is partitioned into channel ``strips.'' For every strip $i$, we compute an importance score $s_i$ and compare it against a global threshold $T$; if $s_i \ge T$ the strip receives 8-bit weights, otherwise 4-bit. Using this bitwidth map, post-training quantization is implemented in PyTorch at the strip level without fine-tuning: only the assigned strips are quantized to their target precision.

Selection of $T$ proceeds via a short sweep guided by the Fisher Information Matrix (FIM). For each candidate $T$, the FIM serves as a sensitivity proxy to estimate accuracy degradation; candidates are ranked jointly by the FIM-predicted accuracy and an energy proxy to identify near-Pareto settings, from which the operating point is chosen.

Hardware cost is obtained on ReRAM crossbars with DNN+NeuroSim. The quantized model is decomposed into 4-bit and 8-bit subsets; each subset is simulated to obtain per-layer energy and latency under the corresponding bitwidth configuration. Results are aggregated according to the observed strip composition to yield end-to-end energy and latency for the mixed-precision deployment.

\begin{table}[h!]
\centering
\caption{Hardware Architecture Configuration}
\begin{tabular}{|p{3.5cm}|p{2cm}|p{2cm}|}
\hline
\textbf{Hardware Architecture} & \multicolumn{2}{c|}{\textbf{Type}} \\ 
\hline
Accelerator Architecture  & \multicolumn{2}{p{4cm}|}{ReRAM } \\ 
\hline
Technology Node    & \multicolumn{2}{p{4cm}|}{32 nm} \\ 
\hline
Array Size    & \multicolumn{2}{p{4cm}|}{128 × 128 synaptic array} \\ 
\hline
Device Precision& \multicolumn{2}{p{4cm}|}{2-bit} \\ 
\hline
Columns of Single ADC & 2 & 4 \\
\hline
Weight Precision  & 4-bit & 8-bit \\
\hline
ADC Resolution  & 16-level & 256-level \\
\hline
\end{tabular}
\end{table}

\subsection{Performance Comparison of Compression Techniques}
We compared our method with Hessian-Aware Pruning (HAP) across key metrics, including accuracy, memory access latency, power consumption, and ADC (Analog-to-Digital Converter) performance on ReRAM Crossbar architectures.

\begin{table}[hbt]
\centering
\caption{Comparison of ResNet20 between HAP and our method.}
\begin{tabular}{|l|c|c|c|c|c|c|}
\hline
\textbf{Method} & \textbf{CR} & \textbf{Acc-top1} & \textbf{Acc-top5} & \textbf{Latency} & \textbf{Energy} \\ \hline
\textbf{HAP} & 74\% & 74.8\% & 98.45\%  & 2.629 ms & 19.42 mJ \\ \hline
\textbf{OURS} & 74\% & 84.63\% & 99.22\% & 1.121 ms & 7.86 mJ  \\ \hline
\end{tabular}
\end{table}

\begin{figure}[htbp]
\centerline{\includegraphics[scale=0.45]{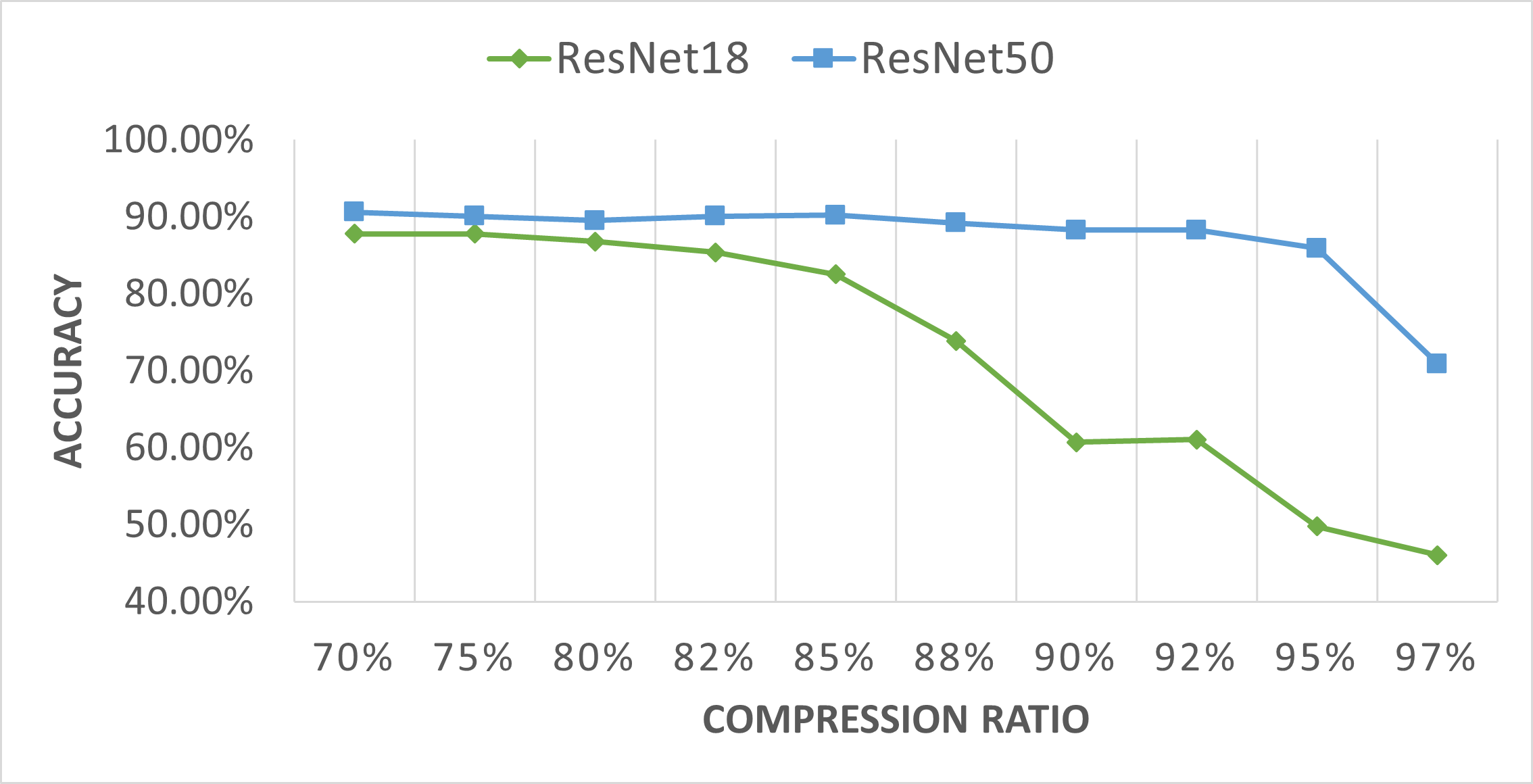}}
\caption{Accuracy Degradation of Different Network Architecture under Increasing Compression Ratios.}
\label{Models_comparison}
\end{figure}

At the same 74\% compression ratio, our method achieved significantly higher accuracy (84.63\% vs. HAP’s 74.8\%), demonstrating better retention of critical weights. This highlights the effectiveness of our intelligent weight selection process.

In terms of efficiency, our method reduced memory access latency by 57\% and power consumption by 60\%, offering substantial improvements for resource-constrained systems, making our approach ideal for energy-sensitive applications like edge computing and embedded systems.

\subsection{Model Complexity versus Quantization Robustness}
\figurename~\ref{Models_comparison} demonstrates that network depth and parameter redundancy strongly affect quantization robustness. ResNet50, with higher representational capacity, shows slower accuracy degradation than ResNet18, as its additional layers and skip connections provide inherent error compensation against precision loss.

From a hardware standpoint, this aligns with the concept of sensitivity heterogeneity, deeper models distribute sensitivity more evenly across layers, allowing the proposed sensitivity-aware mixed-precision quantization to assign lower precision to less critical layers without major accuracy loss.

Both models maintain over 80\% accuracy up to 85\% compression, indicating that many weights can be reduced to 4-bit precision efficiently. Beyond 90\%, however, ResNet18 experiences a sharp decline, highlighting the need for adaptive, layer-specific precision control such as the proposed Dynamic Clustering Optimization.

Overall, deeper models benefit more from mixed-precision schemes, while shallower ones demand finer granularity to preserve accuracy. These findings confirm that model complexity governs quantization resilience and validate the effectiveness of sensitivity-aware quantization for ReRAM-based CIM accelerators.

\subsection{Efficiency Analysis of Mixed-Precision Computation}
We analyzed the performance of mixed-precision compression on ReRAM crossbars using the ResNet18 model and CIFAR-10 dataset. Comparisons were made between 8-bit (no compression), 4-bit (full compression), and mixed-precision configurations (e.g., 10\%, 50\%, 70\% compression ratios) for computational performance and energy consumption.

\begin{table}[h!]
\centering
\caption{Impact of Compression Ratio on Model Accuracy and Energy Breakdown}
\begin{tabular}{|c|c|c|c|c|c|}
\hline
\textbf{CR} & \textbf{Acc} & \textbf{System} & \textbf{ADC} & \textbf{Accumulation} & \textbf{Other} \\ \hline
0\% & 90.91\% & 7.62(mJ) & 7.593(mJ) & 0.67(µJ) & 23.43(µJ) \\ \hline
10\% & 87.25\% & 5.38(mJ) & 5.357(mJ) & 0.63(µJ) & 20.60(µJ) \\ \hline
50\% & 85.58\% & 4.45(mJ) & 4.432(mJ) & 0.58(µJ) & 19.71(µJ) \\ \hline
70\% & 86.33\% & 4.17(mJ) & 4.155(mJ) & 0.57(µJ) & 13.19(µJ) \\ \hline
90\% & 52.01\% & 2.50(mJ) & 2.485(mJ) & 0.49(µJ) & 11.41(µJ) \\ \hline
100\% & 13.88\% & 0.01(mJ) & 0.003(mJ) & 0.23(µJ) & 4.92(µJ) \\ \hline
\end{tabular}
\end{table}

While the 8-bit full-precision model had the highest accuracy (90.91\%), it consumed significantly more power, particularly in the ADC component. Mixed-precision techniques mitigated this, reducing energy use as compression increased. For example, a 70\% compressed model maintained 86.33\% accuracy while reducing power and ADC energy consumption by 40\%.

This demonstrates the balance mixed-precision provides, flexibly mapping weights to lower-bit crossbars, reducing power usage, and optimizing performance.

\subsection{Experimental Evaluation of Crossbar Utilization Optimization}

Table 4 summarizes the impact of the proposed mapping optimization on crossbar utilization.
For ResNet50 (80\% compression, 8-bit quantization), the baseline method achieves only 43.55\% utilization on a 128×128 crossbar, while our approach improves it to 84.36\%.
On a smaller 32×32 crossbar, utilization also rises from 65.92\% to 84.96\%.

\begin{table}[h!]
\scriptsize
\centering
\caption{Bit Utilization Comparison on ResNet50 (80\% Compression Rate)}
\begin{tabular}{lccccc}
\hline
\textbf{Model/CR} & \textbf{Method} & \textbf{Size} & \textbf{Bit} & \textbf{Utilization (\%)} & \textbf{Improvement (\%)}\\ 
\hline
ResNet50/80\% & ORIGIN  & 128$\times$128 & 8bit & 43.55 & - \\
ResNet50/80\% & OUR     & 128$\times$128 & 8bit & 84.36 & +40.81 \\
\hline
ResNet50/80\% & ORIGIN  & 32$\times$32   & 8bit & 65.92 & - \\
ResNet50/80\% & OUR     & 32$\times$32   & 8bit & 84.96 
& +19.04 \\
\hline
\end{tabular}
\end{table}

These results demonstrate that the proposed adjustment strategy effectively enhances crossbar resource usage across different array sizes, with greater benefits observed on larger arrays due to improved alignment of high-bit strip-weights.

\section{conclusion}
This paper introduces an Adaptive Sensitivity-Based Quantization method for ReRAM crossbar arrays, aimed at enhancing energy efficiency and computational performance in deep learning models. Weights are analyzed for sensitivity and split into high- and low-sensitivity clusters, applying 8-bit and 4-bit quantization accordingly. This approach balances computational load and energy savings by mapping high-sensitivity weights to high-precision Crossbars and low-sensitivity weights to low-precision ones.

A dynamic clustering mechanism ensures efficient weight mapping, avoiding issues of unstructured compression. Experiments show that with 70\% compression, the method maintains 86.33\% accuracy while reducing energy consumption by 40\%. This makes it ideal for power-constrained applications like IoT and wearables.

\printbibliography

@INPROCEEDINGS{7551379,
  author={Shafiee, Ali and Nag, Anirban and Muralimanohar, Naveen and Balasubramonian, Rajeev and Strachan, John Paul and Hu, Miao and Williams, R. Stanley and Srikumar, Vivek},
  booktitle={2016 ACM/IEEE 43rd Annual International Symposium on Computer Architecture (ISCA)}, 
  title={ISAAC: A Convolutional Neural Network Accelerator with In-Situ Analog Arithmetic in Crossbars}, 
  year={2016},
  volume={},
  number={},
  pages={14-26},
  doi={10.1109/ISCA.2016.12}}

@INPROCEEDINGS{7551380,
  author={Chi, Ping and Li, Shuangchen and Xu, Cong and Zhang, Tao and Zhao, Jishen and Liu, Yongpan and Wang, Yu and Xie, Yuan},
  booktitle={2016 ACM/IEEE 43rd Annual International Symposium on Computer Architecture (ISCA)}, 
  title={PRIME: A Novel Processing-in-Memory Architecture for Neural Network Computation in ReRAM-Based Main Memory}, 
  year={2016},
  volume={},
  number={},
  pages={27-39},
  doi={10.1109/ISCA.2016.13}}

@INPROCEEDINGS{9218737,
  author={Long, Yun and Lee, Edward and Kim, Daehyun and Mukhopadhyay, Saibal},
  booktitle={2020 57th ACM/IEEE Design Automation Conference (DAC)}, 
  title={Q-PIM: A Genetic Algorithm based Flexible DNN Quantization Method and Application to Processing-In-Memory Platform}, 
  year={2020},
  volume={},
  number={},
  pages={1-6},
  doi={10.1109/DAC18072.2020.9218737}}

@INPROCEEDINGS{9045192,
  author={Sun, Hanbo and Zhu, Zhenhua and Cai, Yi and Chen, Xiaoming and Wang, Yu and Yang, Huazhong},
  booktitle={2020 25th Asia and South Pacific Design Automation Conference (ASP-DAC)}, 
  title={An Energy-Efficient Quantized and Regularized Training Framework For Processing-In-Memory Accelerators}, 
  year={2020},
  volume={},
  number={},
  pages={325-330},
  doi={10.1109/ASP-DAC47756.2020.9045192}}

@INPROCEEDINGS{9218523,
  author={Chu, Chaoqun and Wang, Yanzhi and Zhao, Yilong and Ma, Xiaolong and Ye, Shaokai and Hong, Yunyan and Liang, Xiaoyao and Han, Yinhe and Jiang, Li},
  booktitle={2020 57th ACM/IEEE Design Automation Conference (DAC)}, 
  title={PIM-Prune: Fine-Grain DCNN Pruning for Crossbar-Based Process-In-Memory Architecture}, 
  year={2020},
  volume={},
  number={},
  pages={1-6},
  doi={10.1109/DAC18072.2020.9218523}}

@ARTICLE{9387391,
  author={Meng, Jian and Yang, Li and Peng, Xiaochen and Yu, Shimeng and Fan, Deliang and Seo, Jae-Sun},
  journal={IEEE Transactions on Circuits and Systems II: Express Briefs}, 
  title={Structured Pruning of RRAM Crossbars for Efficient In-Memory Computing Acceleration of Deep Neural Networks}, 
  year={2021},
  volume={68},
  number={5},
  pages={1576-1580},
  doi={10.1109/TCSII.2021.3069011}}

@ARTICLE{10075408,
  author={Lyu, Bo and Wen, Shiping and Yang, Yin and Chang, Xiaojun and Sun, Junwei and Chen, Yiran and Huang, Tingwen},
  journal={IEEE Transactions on Neural Networks and Learning Systems}, 
  title={Designing Efficient Bit-Level Sparsity-Tolerant Memristive Networks}, 
  year={2024},
  volume={35},
  number={9},
  pages={11979-11988},
  doi={10.1109/TNNLS.2023.3250437}}

@ARTICLE{9852786,
  author={Peng, Jie and Liu, Haijun and Zhao, Zhongjin and Li, Zhiwei and Liu, Sen and Li, Qingjiang},
  journal={IEEE Transactions on Computer-Aided Design of Integrated Circuits and Systems}, 
  title={CMQ: Crossbar-Aware Neural Network Mixed-Precision Quantization via Differentiable Architecture Search}, 
  year={2022},
  volume={41},
  number={11},
  pages={4124-4133},
  doi={10.1109/TCAD.2022.3197495}}

@INPROCEEDINGS{9218724,
  author={Qu, Songyun and Li, Bing and Wang, Ying and Xu, Dawen and Zhao, Xiandong and Zhang, Lei},
  booktitle={2020 57th ACM/IEEE Design Automation Conference (DAC)}, 
  title={RaQu: An automatic high-utilization CNN quantization and mapping framework for general-purpose RRAM Accelerator}, 
  year={2020},
  volume={},
  number={},
  pages={1-6},
  doi={10.1109/DAC18072.2020.9218724}}

@ARTICLE{9425549,
  author={Dash, Saurabh and Luo, Yandong and Lu, Anni and Yu, Shimeng and Mukhopadhyay, Saibal},
  journal={IEEE Transactions on Computer-Aided Design of Integrated Circuits and Systems}, 
  title={Robust Processing-In-Memory With Multibit ReRAM Using Hessian-Driven Mixed-Precision Computation}, 
  year={2022},
  volume={41},
  number={4},
  pages={1006-1019},
  doi={10.1109/TCAD.2021.3078408}}

@inproceedings{NIPS1989_6c9882bb,
 author = {LeCun, Yann and Denker, John and Solla, Sara},
 booktitle = {Advances in Neural Information Processing Systems},
 editor = {D. Touretzky},
 pages = {},
 publisher = {Morgan-Kaufmann},
 title = {Optimal Brain Damage},
 volume = {2},
 year = {1989}
}

@article{PAN20141,
title = {Recent progress in resistive random access memories: Materials, switching mechanisms, and performance},
journal = {Materials Science and Engineering: R: Reports},
volume = {83},
pages = {1-59},
year = {2014},
issn = {0927-796X},
doi = {https://doi.org/10.1016/j.mser.2014.06.002},
author = {F. Pan and S. Gao and C. Chen and C. Song and F. Zeng},

}

@article{ZHANG2014415,
title = {A survey of memory architecture for 3D chip multi-processors},
journal = {Microprocessors and Microsystems},
volume = {38},
number = {5},
pages = {415-430},
year = {2014},
issn = {0141-9331},
doi = {https://doi.org/10.1016/j.micpro.2014.03.007},
author = {Yuang Zhang and Li Li and Zhonghai Lu and Axel Jantsch and Minglun Gao and Hongbing Pan and Feng Han},
}

@ARTICLE{6835201,
  author={Kim, Sungho and Zhou, Jiantao and Lu, Wei D.},
  journal={IEEE Transactions on Electron Devices}, 
  title={Crossbar RRAM Arrays: Selector Device Requirements During Write Operation}, 
  year={2014},
  volume={61},
  number={8},
  pages={2820-2826},
  doi={10.1109/TED.2014.2327514}}

@ARTICLE{7312469,
  author={Chen, Pai-Yu and Yu, Shimeng},
  journal={IEEE Transactions on Electron Devices}, 
  title={Compact Modeling of RRAM Devices and Its Applications in 1T1R and 1S1R Array Design}, 
  year={2015},
  volume={62},
  number={12},
  pages={4022-4028},
  doi={10.1109/TED.2015.2492421}}

@BOOK{Sze2020-be,
  title     = "Efficient processing of deep neural networks",
  author    = "Sze, Vivienne and Chen, Yu-Hsin and Yang, Tien-Ju and Emer, Joel
               S",
  publisher = "Springer International Publishing",
  series    = "Synthesis lectures on computer architecture",
  year      =  2020,
  address   = "Cham",
  language  = "en"
}

@INPROCEEDINGS{8702715,
  author={Peng, Xiaochen and Liu, Rui and Yu, Shimeng},
  booktitle={2019 IEEE International Symposium on Circuits and Systems (ISCAS)}, 
  title={Optimizing Weight Mapping and Data Flow for Convolutional Neural Networks on RRAM Based Processing-In-Memory Architecture}, 
  year={2019},
  volume={},
  number={},
  pages={1-5},
  doi={10.1109/ISCAS.2019.8702715}}

@ARTICLE{Gopalakrishnan2020-of,
  title     = "{HFNet}: A {CNN} architecture co-designed for neuromorphic
               hardware with a crossbar array of synapses",
  author    = "Gopalakrishnan, Roshan and Chua, Yansong and Sun, Pengfei and
               Sreejith Kumar, Ashish Jith and Basu, Arindam",
  journal   = "Front. Neurosci.",
  publisher = "Frontiers Media SA",
  volume    =  14,
  pages     = "907",
  month     =  oct,
  year      =  2020,
  copyright = "https://creativecommons.org/licenses/by/4.0/",
  language  = "en"
}

@ARTICLE{9855854,
  author={Zhou, Chuteng and Redondo, Fernando García and Büchel, Julian and Boybat, Irem and Comas, Xavier Timoneda and Nandakumar, S. R. and Das, Shidhartha and Sebastian, Abu and Le Gallo, Manuel and Whatmough, Paul N.},
  journal={IEEE Micro}, 
  title={ML-HW Co-Design of Noise-Robust TinyML Models and Always-On Analog Compute-in-Memory Edge Accelerator}, 
  year={2022},
  volume={42},
  number={6},
  pages={76-87},
  doi={10.1109/MM.2022.3198321}}

@ARTICLE{9965422,
  author={Joardar, Biresh Kumar and Doppa, Janardhan Rao and Li, Hai and Chakrabarty, Krishnendu and Pande, Partha Pratim},
  journal={IEEE Transactions on Emerging Topics in Computing}, 
  title={ReaLPrune: ReRAM Crossbar-Aware Lottery Ticket Pruning for CNNs}, 
  year={2023},
  volume={11},
  number={2},
  pages={303-317},
  doi={10.1109/TETC.2022.3223630}}

@article{10.1109/TPAMI.2022.3195774,
author = {Lin, Mingbao and Zhang, Yuxin and Li, Yuchao and Chen, Bohong and Chao, Fei and Wang, Mengdi and Li, Shen and Tian, Yonghong and Ji, Rongrong},
title = {1xN Pattern for Pruning Convolutional Neural Networks},
year = {2023},
issue_date = {April 2023},
publisher = {IEEE Computer Society},
address = {USA},
volume = {45},
number = {4},
issn = {0162-8828},
doi = {10.1109/TPAMI.2022.3195774},
journal = {IEEE Trans. Pattern Anal. Mach. Intell.},
month = apr,
pages = {3999–4008},
numpages = {10}
}

@inproceedings{10.1145/3489517.3530476,
author = {Negi, Shubham and Chakraborty, Indranil and Ankit, Aayush and Roy, Kaushik},
title = {NAX: neural architecture and memristive xbar based accelerator co-design},
year = {2022},
isbn = {9781450391429},
publisher = {Association for Computing Machinery},
address = {New York, NY, USA},
url = {https://doi.org/10.1145/3489517.3530476},
doi = {10.1145/3489517.3530476},
booktitle = {Proceedings of the 59th ACM/IEEE Design Automation Conference},
pages = {451–456},
numpages = {6},
location = {San Francisco, California},
series = {DAC '22}
}

@INPROCEEDINGS{298572,
  author={Hassibi, B. and Stork, D.G. and Wolff, G.J.},
  booktitle={IEEE International Conference on Neural Networks}, 
  title={Optimal Brain Surgeon and general network pruning}, 
  year={1993},
  volume={},
  number={},
  pages={293-299 vol.1},
  doi={10.1109/ICNN.1993.298572}}

@inproceedings{dong2019hawq,
  title={Hawq: Hessian aware quantization of neural networks with mixed-precision},
  author={Dong, Zhen and Yao, Zhewei and Gholami, Amir and Mahoney, Michael W and Keutzer, Kurt},
  booktitle={Proceedings of the IEEE/CVF international conference on computer vision},
  pages={293--302},
  year={2019}
}

@article{dong2020hawq,
  title={Hawq-v2: Hessian aware trace-weighted quantization of neural networks},
  author={Dong, Zhen and Yao, Zhewei and Arfeen, Daiyaan and Gholami, Amir and Mahoney, Michael W and Keutzer, Kurt},
  journal={Advances in neural information processing systems},
  volume={33},
  pages={18518--18529},
  year={2020}
}

@inproceedings{wang2019eigendamage,
  title={Eigendamage: Structured pruning in the kronecker-factored eigenbasis},
  author={Wang, Chaoqi and Grosse, Roger and Fidler, Sanja and Zhang, Guodong},
  booktitle={International conference on machine learning},
  pages={6566--6575},
  year={2019},
  organization={PMLR}
}

@inproceedings{yu2022hessian,
  title={Hessian-aware pruning and optimal neural implant},
  author={Yu, Shixing and Yao, Zhewei and Gholami, Amir and Dong, Zhen and Kim, Sehoon and Mahoney, Michael W and Keutzer, Kurt},
  booktitle={Proceedings of the IEEE/CVF Winter Conference on Applications of Computer Vision},
  pages={3880--3891},
  year={2022}
}

@inproceedings{karakida2019universal,
  title={Universal statistics of fisher information in deep neural networks: Mean field approach},
  author={Karakida, Ryo and Akaho, Shotaro and Amari, Shun-ichi},
  booktitle={The 22nd International Conference on Artificial Intelligence and Statistics},
  pages={1032--1041},
  year={2019},
  organization={PMLR}
}

@inproceedings{liang2019fisher,
  title={Fisher-rao metric, geometry, and complexity of neural networks},
  author={Liang, Tengyuan and Poggio, Tomaso and Rakhlin, Alexander and Stokes, James},
  booktitle={The 22nd international conference on artificial intelligence and statistics},
  pages={888--896},
  year={2019},
  organization={PMLR}
}

@article{li2018visualizing,
  title={Visualizing the loss landscape of neural nets},
  author={Li, Hao and Xu, Zheng and Taylor, Gavin and Studer, Christoph and Goldstein, Tom},
  journal={Advances in neural information processing systems},
  volume={31},
  year={2018}
}

@inproceedings{chen2017neurosim+,
  title={NeuroSim+: An integrated device-to-algorithm framework for benchmarking synaptic devices and array architectures},
  author={Chen, Pai-Yu and Peng, Xiaochen and Yu, Shimeng},
  booktitle={2017 IEEE International Electron Devices Meeting (IEDM)},
  pages={6--1},
  year={2017},
  organization={IEEE}
}

@inproceedings{peng2019dnn+,
  title={DNN+ NeuroSim: An end-to-end benchmarking framework for compute-in-memory accelerators with versatile device technologies},
  author={Peng, Xiaochen and Huang, Shanshi and Luo, Yandong and Sun, Xiaoyu and Yu, Shimeng},
  booktitle={2019 IEEE international electron devices meeting (IEDM)},
  pages={32--5},
  year={2019},
  organization={IEEE}
}

@article{chen2018neurosim,
  title={NeuroSim: A circuit-level macro model for benchmarking neuro-inspired architectures in online learning},
  author={Chen, Pai-Yu and Peng, Xiaochen and Yu, Shimeng},
  journal={IEEE Transactions on Computer-Aided Design of Integrated Circuits and Systems},
  volume={37},
  number={12},
  pages={3067--3080},
  year={2018},
  publisher={IEEE}
}

@misc{hap_github,
  author       = {Yu, Shixing and Yao, Zhewei and Dong, Zhen and Amir Gholami},
  title        = {Hessian-Aware Pruning (HAP)},
  howpublished = {\url{https://github.com/yaozhewei/HAP}},
  note         = {GitHub repository}
}

\end{document}